\documentclass[global
,twocolumn
]{svjour}
\usepackage{times}
\usepackage{graphicx}
\usepackage{cite}
\journalname{Applied Physics B}

\begin{document}
\title{Time-Resolved Two-Photon Quantum Interference}
\author{T. Legero \and T. Wilk \and A. Kuhn \and G. Rempe}
\institute{Max-Planck-Institut f\"ur Quantenoptik, Hans-Kopfermann-Str.\,1, D-85748 Garching, Germany}
\mail{axel.kuhn@mpq.mpg.de}
\date{Received: \today}
\maketitle
\begin{abstract}
The interference of two independent single-pho\-ton pulses impinging on a beam splitter is analysed in a generalised time-resolved manner. Different aspects of the phenomenon are elaborated using different representations of the single-photon wave packets, like the decomposition into sin\-gle-frequency field modes or  spatio-temporal modes matching the photonic wave packets.  Both representations lead to equivalent results, 
and a photon-by-photon analysis
reveals that the quantum-mechanical two-photon interference can be interpreted as a classical one-photon interference once a first photon is detected. A novel time-dependent  quantum-beat effect is predicted if the interfering photons have different frequencies. The calculation also reveals that full two-photon fringe visibility can be achieved under almost any circumstances by applying a temporal filter to the signal.
\end{abstract}
\section{Introduction}
Single-photon light pulses are a central ingredient of quantum-teleportation and quantum-communication protocols as well as quantum computation with linear optics \cite{Bouwmeester97,Boschi98,Knill01,Bouwmeester00}. Many of these schemes are based on the quantum interference of two indistinguishable photons that impinge simultaneously on different entrance ports of a 50:50 beam splitter.
In this case, both photons leave the beam splitter at the same output port. This quantum interference effect was first observed by  Hong, Ou and Mandel with photons from a parametric down-con\-ver\-sion source \cite{Hong87,Ou88:2,Mandel99}. When the two photons had different frequencies, a spatial quantum-beat phenomenon was observed \cite{Ou88}. The two-photon quantum interference effect has also been observed with photons from independent down-conversion sources \cite{Riedmatten03, Pittman03}, and the phenomenon has been used by Santori et al.\,\cite{Santori02} to demonstrate the indistinguishability of  photons  emitted from a quantum dot embedded in a microcavity. A theoretical description that applies to the interference of independent single photons has been worked out by Bylander et al.\,\cite{Bylander03}.

So far, all these experiments use photons that are short compared to the time resolution of the employed photodetectors, and temporal effects are therefore neglected. However, we have recently developed a novel type of single-photon emitter that employs a strongly coupled atom-cavity system \cite{Kuhn02,Kuhn02:2,Kuhn03} to emit single-photon wave packets of $2\,\mu$s duration. This exceeds the typical response time of solid-state photodetectors by more than three orders of magnitude and allows to study two-photon quantum interference effects in a time-resolved manner. This paper therefore presents a detailed analytical treatment of the two-photon interference process. It takes into account the time evolution of the amplitudes and phases of the interfering single-photon wave packets. The analysis is valid for independent photons, and therefore extends the model for frequency-entangled photon pairs from Steinberg et al. \cite{Steinberg92}.

While preparing this paper, we realised that apparently there is no commonly adopted approach to deal with free-running single-photon wave packets \cite{Mandel95}. One possibility is to express the quantum state of a single-photon wave packet as a superposition state of an infinite number of single-fre\-quency modes, another approach uses a superposition state of spatio-temporal modes. To demonstrate that these approaches lead to the same result,  we now analyse the two-photon interference effect along both routes.

\begin{figure}
\centering\includegraphics[width=8.0cm]{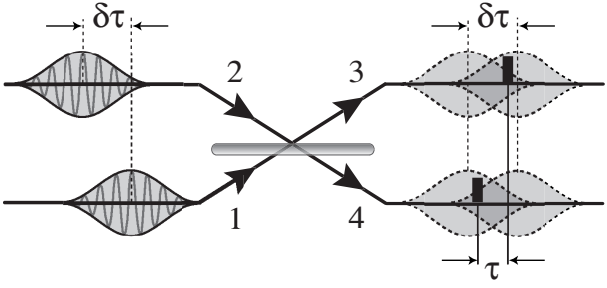}
\caption{\small Two single-photon wave packets impinge with a relative delay of $\delta\tau$ on ports `1' and `2' of a beam splitter. The pulse duration is longer than the time resolution of the detectors monitoring ports `3' and `4' of the beam splitter. Therefore, the time difference, $\tau$, between two photon detections must be taken into account in the analysis of the photon statistics.  \label{beam splitter}}
\end{figure}

\section{Two-photon interference}
As a starting point for the following discussion, we first review the two-photon interference phenomenon in the Fock-state picture, i.e. we consider two single photons that impinge on the entrance ports labelled `1' and `2' of the beam splitter shown in Fig.\,\ref{beam splitter}. This initial situation corresponds to the quantum state $|1_{1}1_{2}\rangle=a^\dag_{1}a^\dag_{2}|0\rangle$, which is obtained by applying the appropriate photon-creation operators $a^\dag_{i}$ on the vacuum state $|0\rangle$. The effect of the 50:50  beam splitter on the field is usually described by the unitary transformations,
\begin{equation}
\begin{array}{ll}
a^\dag_{1}=(a^\dag_{3}+a^\dag_{4})/\sqrt{2}\\
a^\dag_{2}=(a^\dag_{3}-a^\dag_{4})/\sqrt{2}
\end{array}
\quad
\mathrm{and}
\quad
\begin{array}{ll}
a^\dag_{3}=(a^\dag_{1}+a^\dag_{2})/\sqrt{2}\\
a^\dag_{4}=(a^\dag_{1}-a^\dag_{2})/\sqrt{2}
\end{array}.
\label{BSOP}
\end{equation}
If we use these relations to express the initial state in terms of photons created in the two output ports labelled `3' and `4', we immediately see that this leads to an entangled state with either both photons in one or the other output port,
\begin{equation}
\begin{array}{rl}
|1_{1}1_{2}\rangle=a^\dag_{1}a^\dag_{2}|0\rangle
&=\frac{1}{2}(a^\dag_{3}+a^\dag_{4})(a^\dag_{3}-a^\dag_{4})|0\rangle\\
&=\frac{1}{2}([a^\dag_{3}]^2-[a^\dag_{4}]^2)|0\rangle\\
&=(|2_{3}0_{4}\rangle-|0_{3}2_{4}\rangle)/\sqrt{2}.
\end{array}
\end{equation}
This simple picture qualitatively explains the observed interference phenomena, but since the photonic modes that belong to the creation operators  are neither specified in space, frequency nor time (albeit they are assumed to be identical for both photons), quantitative results for photon wave packets cannot be obtained from this approach.

\subsection{Space-time domain}
To obtain more insight, we now consider a description of the arriving single-photon wave packets in the space-time domain. The free-running photons are not subject to boundary conditions that restrict the photon creation and annihilation operators, $a_{k}^\dag$ and $a_{k}$, to specific frequency modes. Therefore we  are free to define these operators in such a way that  they create or annihilate photons in arbitrarily chosen modes (labelled by the index $k$) in space and time, which we define by the one-dimensional spatio-temporal mode functions 
\begin{equation}
\zeta_{k}(z,t)=\epsilon_{k}(t-z/c) e^{-i\phi_{k}(t-z/c)}.
\label{tmode}
\end{equation}
Beside an arbitrary phase evolution, $\phi_{k}(t)$, the mode functions incorporate amplitude envelopes, $\epsilon_{k}(t)$, which are assumed to be normalised so that $\int dt |\epsilon_{k}(t)|^2=1$. By placing the beam splitter at $z=0$, we can omit the spatial coordinate, $z$, in the following. Now we assume that the  whole Hilbert space is spanned by an  orthonormal set of spatio-temporal modes, $\zeta_{k}(t)$, so that the electric field operators read 
\begin{equation}
\hat E^+(t)=\sum_{k}\zeta_{k}(t)a_{k}
\qquad\mathrm{and}\qquad
\hat E^-(t)=\sum_{k}\zeta^*_{k}(t)a^\dag_{k}.
\label{FOPS}
\end{equation}
Provided a photon is present in mode $i$, the probability to measure it at a given time $t$ then is
\begin{equation}
P_{i}(t)=\langle 1_{i}|\hat E^-(t)\hat E^+(t)|1_{i}\rangle
=\zeta^*_{i}(t)\zeta_{i}(t)=|\epsilon_{i}(t)|^2,
\label{piprob}
\end{equation}
which is solely defined by the amplitude envelope of the photon, $\epsilon_{i}(t)$. To analyse the effect of the beam splitter, we now attribute  field operators, $\hat E_{1,2,3,4}^\pm$, to each of the four in\-put/out\-put ports. For the two input ports we only consider the occupied modes which are described by the mode functions $\zeta_{1}(t)$ and $\zeta_{2}(t)$, respectively. This restricts the Hilbert subspaces of the input ports to single modes. Therefore the field operators that belong to the two input ports can be written as
 $\hat E_1^+=\zeta_{1}(t)a_{1}$ and  $\hat E_2^+=\zeta_{2}(t)a_{2}$, respectively. Using these operators, the effect of the beam splitter is now described as a transformation between input and output modes:
\begin{equation}
\begin{array}{rl}
\hat E_3^+(t)&=[\hat E_1^+(t)+\hat E_2^+(t)]/\sqrt{2}\\
&=[\zeta_{1}(t)a_{1}+\zeta_{2}(t)a_{2}]/\sqrt{2}\\[2mm]
\hat E_4^+(t)&=[\hat E_1^+(t)-\hat E_2^+(t)]/\sqrt{2}\\
&=[\zeta_{1}(t)a_{1}-\zeta_{2}(t)a_{2}]/\sqrt{2}
\end{array}
\label{BSField}
\end{equation}
Note that this is a more general form than the simple operator relation Eq.\,(\ref{BSOP}), which is equivalent to  Eq.\,(\ref{BSField})  if all mode functions, $\zeta_{i}$,  are equal. 

\subsubsection{Joint photon-detection probability:}
\label{OpBased}
To obtain the     probability for photon detections in the output ports `3' and `4' at times $t_{0}$ and $t_{0}+\tau$, respectively, we have to apply the field operators $\hat E^\pm_{3}(t_{0})$ and $\hat E^\pm_{4}(t_{0}+\tau)$. With two photons impinging on the two input ports in modes $\zeta_{1}(t)$ and $\zeta_{2}(t)$, the incoming state is $|\Psi_{in}\rangle=a_{1}^\dag a_{2}^\dag |0\rangle$, and the joint probability for photon detections reads
\begin{equation}
\begin{array}{l}
P_{joint}(t_{0},\tau)=g_{34}(t_{0},t_{0}+\tau)=\\
\langle 0|a_{1}a_{2} \hat E^-_{3}(t_{0})\hat E^-_{4}(t_{0}+\tau) \hat E^+_{4}(t_{0}+\tau) \hat E^+_{3}(t_{0}) a_{1}^\dag a_{2}^\dag |0\rangle.
\end{array}
\label{Pcomb2}
\end{equation}
With the field-operator relations defined in Eq.\,(\ref{BSField}) and the restriction to single-photon states, i.e.\ using $\langle 1|a^\dag a^\dag = a a|1\rangle=0$, one easily verifies that Eq.\,(\ref{Pcomb2}) leads to
\begin{equation}
P_{joint}(t_{0},\tau)=\frac{1}{4}|\zeta_{1}(t_{0}+\tau)\zeta_{2}(t_{0})-\zeta_{2}(t_{0}+\tau)\zeta_{1}(t_{0})|^2.
\label{Pcomb}
\end{equation}
The most striking feature of this joint probability function is that it vanishes for $\tau=0$, no matter how different the mode functions of the two incoming photons are. Moreover, if the phase relation between $\zeta_{1}$ and $\zeta_{2}$ is lost for $\tau\gg\tau_{c}$, where $\tau_{c}$ is the mutual coherence time of the incoming photons, the interference term in Eq.\,(\ref{Pcomb}),
averaged over an ensemble of different photon pairs, is zero, and $P_{joint}$ reduces to 
\begin{equation}
\overline{P_{joint}}(t_{0},\tau)\stackrel{\tau\gg\tau_{c}}{\longrightarrow}\frac{1}{4}\left(P_{1}(t_{0})P_{2}(t_{0}+\tau)+P_{2}(t_{0})P_{1}(t_{0}+\tau)\right).
\end{equation}

\subsubsection{Photon-by-photon analysis:}
\label{PbyP}
Despite the fact that the full interference phenomenon is described by Eq.\,(\ref{Pcomb}), it is hard to acquire an intuitive understanding of the interference process from the two-photon approach. This is not the case if one analyses the effect step-by-step. We therefore  use the same spatio-temporal mode functions as before and ask now for the probability to detect a photon  in output port `4' at time $t_{0}+\tau$ conditioned on a photon detection in output port `3' at time $t_{0}$. The probability to find a photon in port `3' reads
\begin{equation}
\begin{array}{l}
P_{3}(t_{0})=
\langle\Psi_{in}|\hat E_{3}^-(t_{0})\hat E_{3}^+(t_{0})|\Psi_{in}\rangle =\\[2mm]
\langle 1_{1}1_{2}|(\hat E_{1}^-(t_{0})+\hat E_{2}^-(t_{0}))\frac{1}{2}(\hat E_{1}^+(t_{0})+\hat E_{2}^+(t_{0}))|1_{1}1_{2}\rangle =\\[2mm]
\displaystyle
\frac{1}{2}(|\epsilon_{1}(t_{0})|^2+|\epsilon_{2}(t_{0})|^2).
\end{array}
\label{P3}
\end{equation}
Note that this expression includes no interference term, since the initial product state is composed of  single-photon Fock states that have no relative phase. The state $|\Psi_{cond}\rangle$, conditioned on the detection of a photon  in port `3' at time $t_{0}$, is obtained by applying the field operator $\hat E_{3}^+(t_{0})$   to the incoming state, $|1_{1}1_{2}\rangle$. After re-normalisation, this leads to
\begin{equation}
\hat E_{3}^+(t_{0})|1_{1}1_{2}\rangle
\stackrel{norm}{\longrightarrow}
\frac{\zeta_{2}(t_{0})|1_{1}0_{2}\rangle+\zeta_{1}(t_{0})|0_{1}1_{2}\rangle}{\sqrt{|\epsilon_{1}(t_{0})|^2+|\epsilon_{2}(t_{0})|^2}}=
|\Psi_{cond}\rangle.
\end{equation}
The conditioned state,  $|\Psi_{cond}\rangle$, describes a single photon that either  impinges on port `1' or port `2' of the beam splitter, with the amplitude and phase relations between these two modes given by $\zeta_{2}(t_{0})$ and $\zeta_{1}(t_{0})$. This photon  now gives rise to classical interference fringes that start in phase at time $t_{0}$, i.e. the second photon is found in the same port  as the first photon if it is detected at the same instant. However, the phase evolves in time according to $\zeta_{1}(t)$ and $\zeta_{2}(t)$. Therefore the probabilities to detect the second photon in output ports `3' or `4' at time $t_{0}+\tau$ read
\begin{equation}
\begin{array}{rl}
P_{3,4}(t_{0}+\tau)&=
\langle\Psi_{cond}|\hat E_{3,4}^-(t_{0}+\tau)\hat E_{3,4}^+(t_{0}+\tau)|\Psi_{cond}\rangle\\[2mm]
&=\displaystyle
\frac{|\zeta_{1}(t_{0}+\tau)\zeta_{2}(t_{0})\pm
\zeta_{2}(t_{0}+\tau)\zeta_{1}(t_{0})|^2}{2(|\epsilon_{1}(t_{0})|^2+|\epsilon_{2}(t_{0})|^2)}.
\end{array}
\label{P4}
\end{equation}
Together with Eq.\,(\ref{P3}), the joint photon-detection probability for a first detection in output port `3' at time $t_{0}$  and a subsequent photon detection in output port `4' at time $t_{0}+\tau$  is
\begin{equation}
\begin{array}{rl}
P_{joint}(t_{0},\tau)&=P_{3}(t_{0})P_{4}(t_{0}+\tau)\\
&=\frac{1}{4}|\zeta_{1}(t_{0}+\tau)\zeta_{2}(t_{0})-\zeta_{2}(t_{0}+\tau)\zeta_{1}(t_{0})|^2.
\end{array}
\label{Pcomb1}
\end{equation}
Result (\ref{Pcomb1}) is identical to expression (\ref{Pcomb}). However, the step-by-step analysis reveals the equivalence between classical     one-photon interference and two-photon interference conditioned on the detection of a first photon. Note that  the field operators at the output-ports commute, i.e. $[\hat E_{3}^+(t_{0}),\hat E_{4}^+(t_{0}+\tau)]=0$.  Therefore the sequence of operators is irrelevant, and for negative detection-time  delay, $\tau$, which corresponds to a reverse order of photon detections, the same results are obtained.

\subsection{Single-frequency field modes}
A standard way to attack the problem of two interfering photons is based on the decomposition of single-photon wave packets into an infinite number of single-frequency field     modes. To do so, a single photon  in one of the two
spatio-temporal input modes $(i=1,2)$ is expressed as one quantum of excitation occupying a superposition of all
possible frequency modes,
\begin{equation}
|1_i \rangle = \int d \omega ~ \Phi_i (\omega) ~ a^\dag_i(\omega) ~ |0\rangle,
\label{onei}
\end{equation}
where  the spectral amplitude of the single-photon pulse,     $\Phi_i(\omega)$,  is  assumed to be normalised so that $\int d\omega ~ \Phi^\ast_i(\omega) \Phi_i(\omega)     = 1$. The continuum of spectral fields allows one to express the time-dependent electric field operators as
\begin{equation}
\begin{array}{ll}
\displaystyle
\hat{E}^+_i(t) =\frac{1}{\sqrt{2\pi}}\int\!\!d\omega ~ e^{-i \omega t} a_i(\omega) 
&~~\mathrm{and}\\[2mm]
\displaystyle
\hat{E}^-_i(t) =\frac{1}{\sqrt{2\pi}}\int\!\!d\omega ~ e^{i \omega t} a^\dag_i(\omega). 
\end{array}
\end{equation}
To evaluate the effect of the field operator, $\hat{E}^+_i(t)$, acting on the single-photon state $|1_{i}\rangle$ defined in Eq.\,(\ref{onei}), we make use of the fact that the Fourier transformation of the spectrum, $\Phi_i(\omega)$, equals the spatio-temporal mode-function,
\begin{equation}
\zeta_i(t)  = \frac{1}{\sqrt{2\pi}} \int d\omega ~ \Phi_i(\omega) ~ e^{-i \omega t},  
\label{Fourier}
\end{equation}
which leads to
\begin{equation}
\begin{array}{rl}
\hat{E}^+_{i}(t) | 1_i \rangle 
&= \frac{1}{\sqrt{2\pi}}\int\!\!\!\int d\omega ~ d\tilde{\omega} ~  e^{-i
  \omega t} a_{i}(\omega)  ~ \Phi_i(\tilde{\omega}) a^\dag_i(\tilde{\omega}) ~ |0\rangle \\[2mm]
&= \frac{1}{\sqrt{2\pi}}\int d\omega ~ \Phi_i(\omega) ~ e^{-i \omega t} |0\rangle \equiv \zeta_i(t) ~ |0\rangle.
\end{array}
\end{equation} 
Therefore, one easily verifies that the  probability to find the photon at time $t$ reads
\begin{equation}
\begin{array}{rl}
P_{i}(t) &= \langle 1_{i}|\hat{E}^-_{i}(t) \hat{E}^+_{i}(t) | 1_i \rangle\\[2mm]
&= \langle 0 |\zeta^\ast_i(t) ~ \zeta_i(t)| 0 \rangle =
\zeta^\ast_i(t)  \zeta_i(t),
\end{array}
\end{equation} 
which is equivalent to  Eq.\,(\ref{piprob}), since the two approaches were mapped onto one another (see Eq.\,(\ref{Fourier})).

In the field-mode picture, the effect of the beam splitter is described by the transformation relations in Eq.\,(\ref{BSOP}), and the quantum state of the two incoming photons, $|1_{1}1_{2}\rangle$, is expressed as a product state of field-mode superposition states, as they are defined in Eq.\,(\ref{onei}). Therefore we make use of the relations in Eq.\,(\ref{BSOP}) and replace 
 $a^\dag_1(\omega_{1})$ and $a^\dag_2(\omega_{2})$ by the respective sums and differences of $a^\dag_3(\omega)$ and $a^\dag_4(\omega)$, which leads to
\begin{equation}
\begin{array}{rll}
|1_{1}1_{2}\rangle 
&=\int\!\!\!\int &d \omega_1 d \omega_2 ~ \Phi_1 (\omega_1)
 \Phi_2 (\omega_2) ~ a^\dag_1 (\omega_1) a^\dag_2
 (\omega_2) |0\rangle\\[2mm]
&=\int\!\!\!\int &d \omega_1 d \omega_2 ~ \Phi_1 (\omega_1)
 \Phi_2 (\omega_2) \\
 &&\times \frac{1}{2}[a^\dag_3 (\omega_1) + a^\dag_4
 (\omega_1)][a^\dag_3 (\omega_2) - a^\dag_4 (\omega_2)]  |0\rangle.
 \end{array}
 \label{one1one2}
\end{equation}
We now  use  Eq.\,(\ref{Pcomb2}) to calculate the joint detection-probability,
\begin{equation}
\begin{array}{l}
P_{joint}(t_0, \tau) =\\
\langle 1_{1}1_{2}| \hat{E}^-_3 (t_0) \hat{E}^-_4 (t_0+\tau) \hat{E}^+_4 (t_0+\tau)
\hat{E}^+_3 (t_0) |1_{2}1_{2}\rangle,
\end{array}
\label{PC2}
\end{equation}
for two photons detected in  output ports `3' and `4'  at times $t_0$ and $t_0+\tau$, respectively. We use Eq.\,(\ref{one1one2}) to express the state $|1_{1}1_{2}\rangle$ in terms of the creation operators $a^\dag_{3}$ and $a^\dag_{4}$ to obtain the right-hand-side of the operator-bracket in Eq.\,(\ref{PC2}), 
\begin{equation}
\begin{array}{l}
\hat{E}^+_4 (t_0+\tau) \hat{E}^+_3 (t_0) |1_{1}1_{2}\rangle=\\[2mm]
\displaystyle
\frac{1}{4\pi} \int d \omega_{1...4}
 \Phi_1(\omega_1) \Phi_2(\omega_2) e^{-i\omega_3 t} e^{-i\omega_4 (t+\tau)}\\
\qquad \times [\delta(\omega_3\!-\!\omega_2) \delta(\omega_4\!-\!\omega_1) -
 \delta(\omega_3\!-\!\omega_1) \delta(\omega_4\!-\!\omega_2)] |0\rangle.
\end{array}
\end{equation}
We now make use of the  Fourier-transformation relation between $\Phi_{i}(\omega)$ and $\zeta_{i}(t)$  from Eq.\,(\ref{Fourier})
to evaluate the above  integral, which leads to
\begin{equation}
\begin{array}{l}
\hat{E}^+_4 (t_0+\tau) \hat{E}^+_3 (t_0) |1_{1}1_{2}\rangle=\\[2mm]
\displaystyle
\frac{1}{2} [\zeta_1(t+\tau) \zeta_2(t) - \zeta_1(t) \zeta_2(t+\tau)] |0\rangle,
\end{array}
\end{equation}
so that the joint photon-detection probability finally reads
\begin{equation}
P_{joint}(t_0, \tau)=\frac{1}{4}|\zeta_1(t_0+\tau)\zeta_2(t_0)-\zeta_1(t_0)\zeta_2(t_0+\tau) |^2,
\end{equation} 
which is again equivalent to Eq.\,(\ref{Pcomb}). Therefore we conclude that the different approaches of modelling single-photon wave packets all lead to the same expression for the joint two-photon detection probability, $P_{joint}(t_{0},\tau)$. But note that the standard way of using {\it single-frequency field modes} is not well adapted to the problem and requires a lengthy and cumbersome calculation. 
This is not the case in the {\it space-time domain}, where the joint detection probability can be calculated in a fast and elegant way (section \ref{OpBased}). However, physical insight is only obtained from  the {\it photon-by-photon analysis} (section \ref{PbyP}), which reveals that the two-photon quantum interference is in fact equivalent to a classical interference phenomenon once a first photon detection has taken place. In the collapsed state, the relative phases and amplitudes of the two interfering modes are fixed. Therefore an oscillating quantum-beat signal is expected if the two spatio-temporal mode functions, $\zeta_{1}(t)$ and $\zeta_{2}(t)$, evolve at different rates.

\begin{figure}
\centering\includegraphics[width=7.5cm]{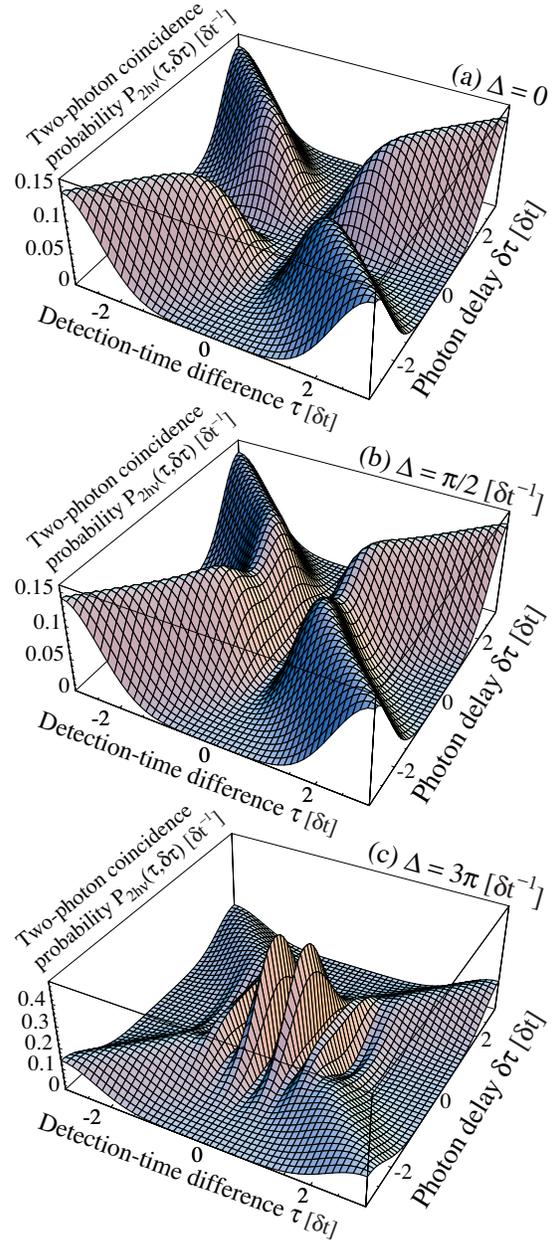}
\caption{\small Joint probability, $P_{2h\nu}$, for the detection of photon pairs as a function of the relative delay between the two impinging photon wave packets, $\delta\tau$, and the delay between photon detections, $\tau$ for the three frequency differences $\Delta=0;\pi/2;3\pi$. All times and frequencies are normalised by the photon-pulse duration, $\delta t$, and, hence, are dimensionless. \label{P2hnu}}
\end{figure}

\section{Quantum beat}
To illustrate the quantum-beat signal in the interference of two photons with slightly different frequency, we now consider the case of two equally long Fourier-limited Gaussian single-photon pulses that impinge on a 50:50 beam splitter with a relative delay time $\delta\tau$ and carrier-frequency difference $\Delta=\omega_{2}-\omega_{1}$. These pulses are described by the two spatio-temporal mode functions
\begin{equation}
\begin{array}{rcl}
\zeta_{1}(t)&=&\sqrt[4]{2/\pi}\exp(-(t-\delta\tau/2)^2-i(\omega-\Delta/2)t)\\
\mathrm{and}&&\\
\zeta_{2}(t)&=&\sqrt[4]{2/\pi}\exp(-(t+\delta\tau/2)^2-i(\omega+\Delta/2)t),
\end{array}
\label{zeta12def}
\end{equation} 
with $\omega=(\omega_{1}+\omega_{2})/2$. For the sake of simplicity, the time, $t$, the photon delay, $\delta\tau$, and the detection-time delay, $\tau$, are expressed  in units of the pulse duration, $\delta t$ (half width at $1/e$ maximum), while the frequencies, $\omega$ and $\Delta$, are expressed  in units of $1/\delta t$. 

With the spatio-temporal modes defined in Eq.\,(\ref{zeta12def}), a     straightforward evaluation of Eq.\,(\ref{Pcomb}) leads to the joint photon-detection probability
\begin{equation}
\begin{array}{l}
P_{joint}(t_{0},\tau,\delta\tau,\Delta)=\\[2mm]
\displaystyle
\frac{\cosh(2 \tau \delta\tau) - \cos(\tau \Delta)}{\pi}\times\exp(-4t_{0}(t_{0}+\tau)-\delta\tau^2-2\tau^2),
\end{array}
\end{equation}
which depends on the photon detection times, $t_{0}$ and $t_{0}+\tau$. To obtain the probability of detecting two photons in the ports `3' and `4' with a time difference $\tau$, we integrate over all possible values of $t_{0}$, which leads to 
\begin{equation}
\begin{array}{l}
P_{2h\nu}(\tau,\delta\tau,\Delta)=g^{(2)}_{3,4}(\tau) |_{\delta\tau,\Delta}=\\[2mm]
\int dt_{0}P_{joint}(t_{0},\tau,\delta\tau,\Delta)=\\[2mm]
\displaystyle
\frac{\cosh(2 \tau \delta\tau) - \cos(\tau \Delta)}{2\sqrt{\pi}}\times\exp(-\delta\tau^2-\tau^2).
\end{array}
\end{equation}
This joint two-photon detection probability is displayed in Fig.\,\ref{P2hnu} as a function of the photon-delay, $\delta\tau$, and the detection-time delay, $\tau$,  for three different values of the frequency difference, $\Delta$.

The case with $\Delta=0$ is depicted in Fig.\,\ref{P2hnu}(a). As expected, no coincidences are found if the photons arrive simultaneously, i.e. for $\delta\tau=0$. This situation  is well explained by the Fock-state model of two interfering indistinguishable photons. Moreover, even if the two photons are delayed with respect to each other,  we see that simultaneous
photon detections (with $\tau=0$) never occur. This is well explained by Eq.\,(\ref{Pcomb}) which yields that $P_{joint}(t_{0},\tau=0)\equiv 0$, no matter how different the mode functions are. In particular, the  second photon 
leaves the beam splitter through the same port as the first one, provided no dephasing takes place, i.e. for
$|\zeta_1(t_0+\tau)\zeta_2(t_0)-\zeta_1(t_0)\zeta_2(t_0+\tau) |\ll 1$.

Figure \ref{P2hnu}(b and c) show situations where the two impinging photon wave packets have a frequency difference of $\Delta=\pi/2$ or $\Delta=3\pi$, respectively. Most remarkable is that simultaneous detections (with $\tau=0$) still do not occur, although the two photons are now distinguishable. Moreover, for simultaneously arriving photon wave packets (with $\delta\tau=0$), the coincidence probability now oscillates as a function of $\tau$ with the frequency difference $\Delta$. We emphasise that this beat note always starts at zero for $\tau=0$, due to the in-phase starting condition imposed by the detection of the first photon. 

\begin{figure}
\centering\includegraphics[width=7.5cm]{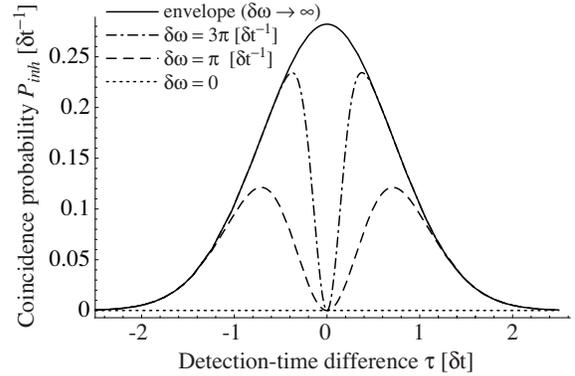}
\caption{\small Time-resolved two-photon detection probability, $P_{inh}$, as a function of the detection-time difference, $\tau$, in case of simultaneously arriving equally long single-photon wave packets $(\delta\tau=0)$ for different  inhomogeneous spectral widths, $\delta\omega$. The solid line represents non-interfering photons and therefore reflects the amplitude envelope of the photonic wave packets. \label{Pbw}}
\end{figure}

It follows that the coincidence probability remains zero even in case of an inhomogeneous broadening of the frequency differences, $\Delta$. To illustrate this, we assume a Gaussian frequency distribution,
\begin{equation}
f_{\delta\omega}(\Delta)=\frac{1}{\delta\omega \sqrt{\pi}}\exp(-(\Delta/\delta\omega)^2),
\end{equation}
where $\delta\omega$ defines the bandwidth of the spectrum (half width at $1/e$ maximum). The integral of this frequency distribution is normalised to one, so that we can use it to calculate the photon detection probability for simultaneously arriving photon wave packets (with $\delta\tau=0$) by evaluating
\begin{equation}
\begin{array}{rl}
\displaystyle
P_{inh}(\tau,\delta\omega)&
\displaystyle
=\int d\Delta ~ f_{\delta\omega}(\Delta)P_{2h\nu}(\tau,\delta\tau=0,\Delta)\\[2mm]
&\displaystyle
=\frac{\exp(-\tau^2)}{2\sqrt{\pi}}\left(1-\exp\left(-\left(\frac{\tau}{2/\delta\omega}\right)^2\right)\right).
\end{array}
\end{equation}
Figure \ref{Pbw} shows that the inhomogeneous broadening of the photonic spectrum leads to a dip in the two-photon detection probability with width $2/\delta\omega$ (half width at 1/e dip-depth). We emphasise that this dip in $P_{inh}$ reaches zero. Therefore it is possible to achieve perfect two-photon coalescence if a temporal filter is applied that restricts the detection-time difference to $|\tau|\ll 2/\delta\omega$.

\begin{figure}
\centering\includegraphics[width=7.5cm]{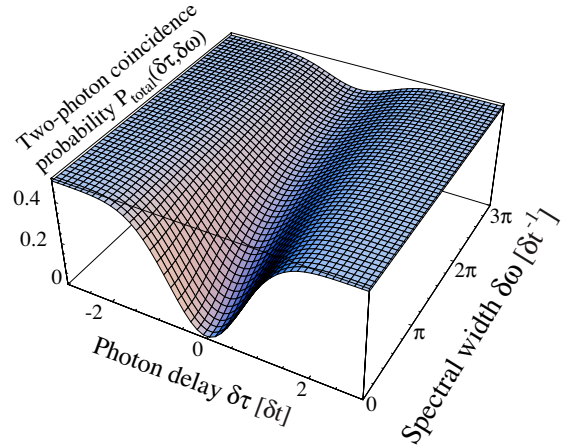}
\caption{\small Total probability for the detection of photon pairs as a function of the relative delay between the two impinging photon wave packets, $\delta\tau$, and the inhomogeneous spectral width, $\delta\omega$, of the interfering photons. \label{Ptotal}}
\end{figure}

To connect our results to previous experiments with ultra-short photons, we now calculate the total two-photon coincidence probability  by integrating over the detection-time delay, $\tau$. In case of inhomogeneous broadening, as discussed above, this leads to
\begin{equation}
\begin{array}{rl}
\displaystyle
P_{total}(\delta\tau,\delta\omega)&
\displaystyle
=\int\!\!\!\!\int d\tau ~ d\Delta ~ f_{\delta\omega}(\Delta)P_{2h\nu}(\tau,\delta\tau,\Delta)\\[3mm]
&\displaystyle
=\frac{1}{2}-\frac{\exp(-\delta\tau^2)}{\sqrt{4+\delta\omega^2}}.
\end{array}
\end{equation}
The total two-photon coincidence probability is shown in     Fig.\,\ref{Ptotal} as a function of the arrival-time delay, $\delta\tau$, and the inhomogeneous spectral width, $\delta\omega$. In this case, it is evident that the two-photon coincidence dip reaches zero only if $\delta\omega=0$,  i.e. only if the two photons are mutually coherent. Increasing the bandwidth has no effect on the width of the dip, but only decreases its depth. Therefore spectral filtering is mandatory if the time resolution does not allow to filter photon-detection times.

\section{Conclusion} 
A rich substructure is expected in two-photon interference experiments if the photons can be detected with high time resolution.
For example, distinguishable photons of different frequencies 
are expected to give rise to a pronounced oscillation of the joint photon-detection probability behind the beam splitter. 
The ability of generating long single-photon pulses by means of a strongly-coupled atom-cavity system opens up the possibility to `zoom' into the quantum state of a single-photon pulse by measuring two-photon quantum interference fringes as a function of time.

\begin{acknowledgement}
This work was supported by the focused research program `Quantum Information Processing' of the Deutsche Forschungsgemeinschaft, and by the European Union through the IST (QUBITS) and IHP (QUEST) programs. We also thank M. Hennrich for many valuable comments.
\end{acknowledgement}

\bibliographystyle{unsrt}

\end{document}